\documentclass[10pt,conference]{IEEEtran}

\usepackage[utf8]{inputenc}
\usepackage{cite}
\usepackage{amsmath,amssymb,amsfonts}
\usepackage{algorithm}
\usepackage{algpseudocode}
\usepackage{graphicx}
\usepackage{textcomp}
\usepackage{xcolor}
\usepackage{listings}
\usepackage{hyperref}
\usepackage{caption}
\usepackage{subcaption}
\usepackage{array}
\usepackage{multirow}
\usepackage[normalem]{ulem}
\usepackage{booktabs}
\usepackage{calc}

\usepackage{comment}

\graphicspath{{figures}}

\newcommand{\cbraces}[1]{\{#1\}}

\newcommand{\approachPipelines}[0]{AutoPABS}

\definecolor{new}{RGB}{0,0,0}
\definecolor{new2}{RGB}{0,0,0}
\definecolor{new3}{RGB}{0,0,0}

\newcommand{\makecellpos}[2]{\begin{tabular}{@{}#1@{}}#2\end{tabular}}
\newcommand{\makecell}[1]{\makecellpos{l}{#1}}

\definecolor{deepred}{rgb}{0.6,0,0}
\definecolor{eclipseStrings}{RGB}{42,0.0,255}
\definecolor{eclipseKeywords}{RGB}{127,0,85}
\colorlet{numb}{magenta!60!black}

\lstdefinelanguage{JSON}{
	basicstyle=\normalfont\ttfamily\footnotesize,
	commentstyle=\color{gray}, 
	stringstyle=\color{eclipseStrings},
	showstringspaces=false,
	breaklines=true,
	string=[s]{"}{"},
	literate=
	*{0}{{{\color{deepred}0}}}{1}
	{1}{{{\color{deepred}1}}}{1}
	{2}{{{\color{deepred}2}}}{1}
	{3}{{{\color{deepred}3}}}{1}
	{4}{{{\color{deepred}4}}}{1}
	{5}{{{\color{deepred}5}}}{1}
	{6}{{{\color{deepred}6}}}{1}
	{7}{{{\color{deepred}7}}}{1}
	{8}{{{\color{deepred}8}}}{1}
	{9}{{{\color{deepred}9}}}{1}
}

\newcommand{\indentSpec}[0]{\hspace*{10pt}}
\newcommand{\indentSpecNewline}[1]{\indentSpec{}\hspace*{\widthof{#1}}}

\usepackage{flushend}

\begin{document}

\title{Automating Pipelines of A/B Tests with Population Split Using Self-Adaptation and Machine Learning}

\author{\IEEEauthorblockN{Federico Quin}
\IEEEauthorblockA{\textit{Department of Computer Science} \\
\textit{KU Leuven}\\
Leuven, Belgium \\
federico.quin@kuleuven.be\vspace{-10pt}}
\and
\IEEEauthorblockN{Danny Weyns}
\IEEEauthorblockA{\textit{Department of Computer Science} \\
\textit{KU Leuven, Belgium} \\
\textit{Linnaeus University, Sweden}\\
danny.weyns@kuleuven.be\vspace{-10pt}}
}

\maketitle

\begin{abstract}
A/B testing is a common approach used in industry to facilitate innovation through the introduction of new features or the modification of existing software. Traditionally, A/B tests are conducted sequentially, with each experiment targeting the entire population of the corresponding application. This approach can be time-consuming and costly, particularly when the experiments are not relevant to the entire population. To tackle these problems, we introduce a new self-adaptive approach called \approachPipelines{}, \textcolor{new}{short for \textbf{Auto}mated \textbf{P}ipelines of \textbf{A/B} tests using \textbf{S}elf-adaptation}, that (1) automates the execution of pipelines of A/B tests, and (2) supports a split of the population in the pipeline to divide the population into multiple A/B tests according to user-based criteria, leveraging machine learning. We started the evaluation with a small survey to probe the appraisal of the notation and infrastructure of \approachPipelines{}. Then we performed a series of tests to measure the gains obtained by applying a population split in an automated A/B testing pipeline, using an extension of the SEAByTE artifact. The survey results show that the participants express the usefulness of automating A/B testing pipelines and population split. The tests show that automatically executing pipelines of A/B tests with a population split accelerates the identification of statistically significant results of the parallel executed experiments of A/B tests compared to a traditional approach that performs the experiments sequentially. 
\end{abstract}

\section{Introduction}

\renewcommand{\arraystretch}{1.2}

A/B testing, also referred to as online controlled experimentation, continuous experimentation, bucket testing, or randomized experimentation, forms a crucial part of modern software businesses such as Google, Amazon, or Meta. A/B testing supports businesses to grow and innovate their customer-facing software applications~\cite{Ha-Thuc2020, Wang2019, Li2015Prediction, Tang2010, Li2019, Xu2016}. The aim of A/B testing is to 
make data-driven decisions to improve the products offered to customers. 
\color{new}
In essence, A/B testing compares two different versions of a software product or service, variant A and variant B, by exposing them to end-users and evaluating the performance of each variant. Unlike traditional software testing methods, A/B testing takes place within live systems and provides real-world data that organizations can use to make well-informed 
decisions~\cite{King2017, Siroker2013, Deng2016}.

\begin{figure}[!bht]
    \centering
    \includegraphics[width=.5\linewidth]{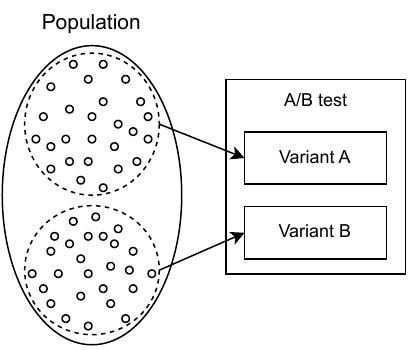}
    \caption{\color{new2}Example of 50/50 population assignment during the execution of an A/B test.}
    \label{fig:intro-population}
\end{figure}

The A/B testing process comprises three phases: design, execution, and analysis~\cite{IssaMattos2018}. 
The design of the A/B test consists of defining key parameters such as the hypothesis to compare the variants, experiment duration, the 
\textcolor{new2}{assignment} of users to both variants and metrics to be collected are defined (designed phase). \textcolor{new2}{We refer to the group of users that take part in the A/B test as the A/B test population.} The metrics, such as click-through rate, number of clicks, and number of sessions~\cite{Bhamidipati2017, Kohavi2013}, are used to quantify the performance of each variant during the experiment.
Once the experiment is designed, both variants are deployed in the live system and the population is split between them (execution phase)\textcolor{new2}{, as shown in Figure~\ref{fig:intro-population}}. The system tracks relevant data during the test, and after the experiment is complete, the hypothesis is tested using a statistical test, e.g. a Student's t-test or Welsh's t-test~\cite{Machmouchi2017, Poyarkov2016, Goswami2015}. The test results provide valuable insights into the performance of each variant, and organizations can use this information to make decisions about which variant to use (analysis phase). 
\color{black}

Traditionally, A/B tests are administered manually, and the results of the tests are analyzed sequentially~\cite{Sajeev2021, Tagami2014}. This process can be costly and time-consuming~\cite{Kohavi2007, Gupta2019, Ros2018, Gardey2020}. Different researchers have therefore argued for further automating the A/B testing process~\cite{Goswami2015, primary-study-12, primary-study-747, Mattos2017, Kumar2019, Gardey2020}. Automating A/B testing is the first challenge we aim to tackle in this paper. Furthermore, conducting A/B tests on the entire population may not be optimal and even result in irrelevant outcomes due to the diverse characteristics of the population~\cite{primary-study-84, Zhao2019}. Although previous work has looked into a segmented analysis of A/B tests~\cite{Zhao2017}, the execution of the A/B tests themselves do not target particular segments of the population. Yet, targeting specific segments of the population for A/B testing can increase the efficiency of the analysis of A/B test results. Targeting A/B tests to specific segments of their user base is the second challenge we aim to tackle in this paper. These challenges are also confirmed by a recent systematic literature review we performed on A/B testing~\cite{Quin2023SLRAB}. 
This brings us to the following problem statement we tackle in this paper:

\begin{quote}
    \textit{How can we automate A/B testing pipelines and run them efficiently?}
\end{quote}

To address this problem, we propose a self-adaptive solution called \approachPipelines{} (Automated Pipelines of A/B tests using Self-adaptation) that handles the deployment, monitoring, analysis, \textcolor{new}{and execution} of pipelines of A/B tests automatically. 
\approachPipelines{} interprets the outcome of A/B tests to initiate subsequent tests specified in the pipeline. To enhance efficiency, we introduce population splits to A/B testing pipelines to target A/B tests to specific segments of the population. We focus on segmenting a population based on the properties or behaviors of users, leveraging machine learning. Segmenting the population enables multiple parts of an A/B testing pipeline to be executed in parallel improving the efficiency of the test execution. We evaluate the usefulness of \approachPipelines{} with a small survey and test the gains of a population split for an online web-store application. 

\textcolor{new3}{In contrast to most research on self-adaptation that aim at novel approaches for engineering self-adaptive systems, \approachPipelines{} takes a complementary angle and aims at supporting a key task of software engineers using self-adaptation, in particular
enhancing the automation of executing A/B testing pipelines. This aligns with recent initiatives, such as Self-Adaptation
2.0~\cite{Bures2021} that argues for an equal-to-equal relationship between
self-adaptation and AI, benefiting one another, and self-adaptation that is applied to deal with
degraded machine-learning components to maintain system utility~\cite{Casimiro2021}.} 

\color{new2} This paper presents the following three contributions:
\begin{itemize}
    \item A specification and notation for A/B testing pipelines
    \item A self-adaptive architecture that enables automated execution of A/B testing pipelines
    \item A population split component that enables more efficient A/B pipeline execution
\end{itemize}
\color{black}

The paper is structured as follows. Section~\ref{sec:related-work} describes related work on automating A/B testing and the use of machine learning in A/B testing. 
In Section~\ref{sec:approach}, we present \approachPipelines{}, the new approach for automating pipelines of A/B tests using self-adaptation and we introduce population splits in pipelines. In Section~\ref{sec:evaluation}, we present the evaluation of \approachPipelines{}. 
Finally, we wrap up and look at future research directions in Section~\ref{sec:conclusion}.

\section{Related Work}\label{sec:related-work}

We discuss a selection of related work for the two main lines of related research: the automation of A/B testing pipelines and the use of machine learning in A/B tests. Then, we position our work in the current landscape of research.
\vspace{3pt}\\\textbf{Automation of A/B testing pipelines.}
\textcolor{new}{Automation of A/B testing has received limited attention in the literature.}
Tamburelli et al.~\cite{Tamburrelli2014} approach A/B testing as an optimization problem that is solved using automated search. Developers annotate program features and the framework automatically generates, selects, and enacts A/B variants. Mattos et al.~\cite{Mattos2017} put forward an architecture framework to model automated experimentation in software systems that they briefly evaluate in a human-robot context. \textcolor{new}{Fabijan et al.~\cite{Fabijan2021} describe an iterative software engineering process to accelerate the use of A/B testing from experience at Microsoft, Outreach, and Booking.com, lowering the human cost of A/B testing and accelerating innovation.}
Researchers have also studied A/B tests and automated deployment based on principles of workflow and task orchestration. R\'ev\'esz et al.~\cite{Revesz2022} target long-term A/B tests and automated deployment in the context of CI/CD, leveraging container orchestration systems to realize the approach.
In an alternative setting, a challenging issue concerning automating A/B tests involves the identification of machine learning models that achieve satisfactory results in a live context~\cite{Treveil2020} (\textcolor{new}{as opposed to offline evaluation 
based on historical data~\cite{Bottou2013, Li2021, Gruson2019}}). To that end, Dai et al.~\cite{Dai2020} present an approach that automatically selects machine learning models to A/B test in live systems, giving priority to promising models.
Another perspective of automating A/B testing pipelines is the gradual roll-out of software releases. Schermann et al.~\cite{Schermann2016} present a modeling approach that supports a gradual roll-out of live testing of a system by setting up multiple sequential A/B tests. The Follow-The-Best-Interval algorithm proposed by Munoz et al.~\cite{Munoz2018} handles the roll-out process automatically.
Gerostathopoulos et al.~\cite{Gerostathopoulos2018} present a tool for end-to-end optimization of a target system, providing a basis for a system to self-optimize through automated experimentation.

Related industrial efforts include Feature Flags and Argo Rollouts. Feature Flags~\cite{Feature-Flags} are if/else controls in a code base. This industrial approach enables faster and safer development, making it easy to manage features without pushing a change by separating deployment from release. 
Argo Rollouts~\cite{Argo-Rollouts} enables a user to run two versions of an application for a specific duration and perform an analysis of their application, for instance, start a baseline and canary deployment in parallel, and compare the metrics produced by the two. 
\vspace{3pt}\\\textbf{Use of machine learning in A/B tests.}
Several researchers have used machine learning to improve the execution of A/B tests. One such case is learning sensitive metric combinations in A/B testing~\cite{Kharitonov2017}. Other work looks at increasing the sensitivity in A/B testing, yielding more reliable and faster outcomes.
Guo et al.~\cite{Guo2021} and Syrgkanis et al.~\cite{Syrgkanis2019} use linear regression predictions of experiment outcomes alongside variance estimators to improve variance reduction in A/B testing, leading to more precise inferences with less data. Poyarkov et al.~\cite{Poyarkov2016} propose a similar approach using boosted decision tree regression.
From a different angle, Li et al.~\cite{Li2015Prediction} make use of diversified historical data and machine learning to make predictions on the A/B metrics of A/B tests, but without running the tests on live systems. \textcolor{new}{Conversely, Zhao et al.~\cite{Zhao2017} employ unsupervised learning techniques in the analysis phase of A/B testing to classify users based on their behavior and analyze test results accordingly.}%
\vspace{3pt}\\\textbf{Positioning and Challenges Tackled.}
The state-of-the-art in A/B testing points to the labor-intensiveness of setting up, analyzing, and conducting A/B tests. Hence an important challenge is further automation of A/B testing. In addition, current research highlights that A/B tests are costly to run in live software systems (A/B tests are typically run for a long time to obtain ample observations). This underpins the challenge of running A/B tests more efficiently, for instance by using machine learning to improve the sensitivity of the A/B tests. Our work aims at contributing to these two challenges, on the one hand by exploiting self-adaptation as a means to automate the execution of A/B testing pipelines and on the other hand through support for splitting populations to focus testing leveraging machine learning.    

\section{Approach}\label{sec:approach}

We now present \approachPipelines{}, the new approach to automate pipelines of A/B tests with support for population splits. We start with outlining the requirements for a solution. Then, we explain how we automate the execution of pipelines of A/B tests using self-adaptation. Next, we zoom in on splitting a population in a pipeline leveraging machine learning. Lastly, we present a concrete implementation of \approachPipelines{}. 

\subsection{Requirements}

The requirements for a solution to automate the execution of pipelines of A/B tests with support for population split are: 

\begin{enumerate}
    \item[\textbf{R1}] To provide a specification for modeling pipelines of A/B tests; 
    \item[\textbf{R2}] To design a conceptual architecture that automates the execution of the modeled pipelines of A/B tests;  
    \item[\textbf{R3}] To provide a specification for modeling a population split in a pipeline of A/B tests; 
    \item[\textbf{R4}] To design an extended conceptual architecture to support population splits when executing a pipeline of A/B tests.     
\end{enumerate}

As an additional requirement, \textbf{R5}, an infrastructure is required that implements the extended conceptual architecture. 

\subsection{Self-adaptation to Automate A/B Testing Pipelines}

\approachPipelines{} leverages the principles of self-adaptation~\cite{Roadmap2009, DeLemos2013, DeLemos2017, Weyns2021introduction} to automate the execution of pipelines of A/B tests. \approachPipelines{} adds a feedback loop~\cite{Kephart2003, Weyns2013External} (managing system) on top of a running system (managed system) that is responsible to deploy and execute a pipeline of A/B tests. \approachPipelines{} assumes that the managed system is ''A/B testing-enabled'' meaning, i.e., it is endowed with capabilities to deploy, monitor, and run A/B tests during operation. 

\begin{figure}
    \centering
    \includegraphics[width=\linewidth]{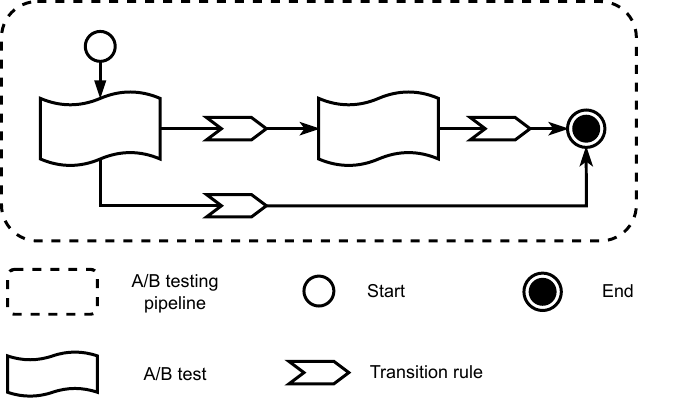}
    \caption{Visual notation of an A/B testing pipeline}
    \label{fig:notation-visual-pipeline}\vspace{-10pt}
\end{figure}

We explain self-adaptation in \approachPipelines{} in two steps. First, we present a specification to model pipelines of A/B tests (tackling requirement R1). Second, we present the conceptual architecture of \approachPipelines{} focusing on the automated execution of pipelines of A/B tests (tackling requirement R2).  
\vspace{5pt}\\\textbf{Specification to model pipelines of A/B tests.}
To support modeling an A/B testing pipeline, we put forward a simple specification. Figure~\ref{fig:notation-visual-pipeline} shows a visual notation of the different elements for a basic example of an A/B testing pipeline. 
Subsequently, we explain the specification of an A/B test, transition rules, and an A/B testing pipeline.

\vspace{3pt}\paragraph{Specification A/B test} \mbox{}\vspace{5pt} \\
\indentSpec{}\textit{AB-test = $<$ Exp-length, AB-assignment, Hypothesis, \newline 
\indentSpecNewline{\textit{AB-test = $<$ }}\{AB-metrics\}, Stat-test $>$}\\\vspace{-5pt}

The experiment length of an A/B test (\textit{Exp-length}) denotes the number of observations or the duration of the test required to complete the experiment. The A/B assignment (\textit{AB-assignment}) specifies the proportions of the population that use variant A and variant B, respectively. The hypothesis (\textit{Hypothesis}) is the supposition that is put forward before the A/B test is conducted. The A/B metrics (\textit{\{AB-metrics\}}) are quantifiable measures used to validate the hypothesis. The statistical test (\textit{Stat-test}) is used to test the hypothesis once the A/B test finishes (due to the number of observations or duration). 

\paragraph{Specification transition rule} \mbox{}\vspace{5pt}\\
\indentSpec{}\textit{Trans-rule = $<$ Assoc-AB-test, Cond-stat, Subseq-AB-test $>$}\\\vspace{-5pt}

A transition rule applies to an A/B test, i.e., the associated A/B test (\textit{Assoc-AB-test}) or the \textit{End} element that ends the pipeline (see below). The conditional statement (\textit{Cond-stat}) is a boolean expression on the outcome of the associated A/B test. The subsequent A/B test (\textit{Subseq-AB-test}) is the next A/B test in the pipeline that should be started if the conditional statement is satisfied. Algorithm~\ref{alg:application-transition-rule} describes the semantics of a transition rule given an A/B test and the result of the test.

\begin{algorithm}
\caption{Test the application of a transition rule.}\label{alg:application-transition-rule}
\begin{algorithmic}[1]
\Procedure{rule-applies}{$\text{trans-rule}, \text{result}, \text{AB-test}$}
\State \Return
\State \quad \quad $\text{AB-test} = \text{trans-rule.Assoc-AB-test}$ \text{\textbf{and}}
\State \quad \quad $\text{trans-rule.Cond-stat.test(result)} = \text{\textbf{true}}$
\EndProcedure
\end{algorithmic}
\end{algorithm}

\paragraph{Specification of A/B testing pipeline} \mbox{}\vspace{5pt}\\
\indentSpec{}\textit{AB-test-pl = $<$ \{AB-test\}, \{Trans-rule\}, Start, End $>$}\\\vspace{-5pt}

An A/B testing pipeline consists of a set of A/B tests (\textit{\{AB-test\}}) and a set of transition rules (\textit{\{Trans-rule\}}). The start element (\textit{Start}) points to the first A/B test in the pipeline, and the end element (\textit{End}) marks the end of the pipeline execution. The end element is of the type A/B test.
The set of transition rules defines the execution flow of the pipeline, based on the results of the executed A/B tests. It is the responsibility of the designer to create consistent transition rules that ensure a proper execution of the pipeline. 
Algorithm~\ref{alg:automate-ab-pipeline} describes the semantics of the automatic execution of A/B testing pipelines.

\begin{algorithm}
\caption{Automatically execute A/B testing pipelines.}\label{alg:automate-ab-pipeline}
\begin{algorithmic}[1]
\Procedure{ExecutePipeline}{pipeline}
\State $\text{test} \gets \text{pipeline.Start}$ \label{algo:pipeline1}
\While{$\text{test} \neq \text{End}$} \label{algo:pipeline2}
\State $\text{res} \gets \text{Deploy(test)}$ \label{algo:pipeline3}
\State $\text{next} \gets \text{End}$
\For{$\text{rule}\ \text{in}\ \text{pipeline.Trans-rule}$} \label{algo:pipeline4}
    \If {Rule-Applies(rule, res, test)}
        \State $\text{next} \gets \text{rule.Subseq-AB-test}$
        \State \textbf{break} \label{algo:pipeline5}
    \EndIf
\EndFor
\State $\text{test} \gets \text{next}$ \label{algo:pipeline6}
\EndWhile
\EndProcedure
\end{algorithmic}
\end{algorithm}

The execution of the A/B testing pipelines starts with the initial A/B test on line~\ref{algo:pipeline1}. Until the End of the pipeline is not encountered (line~\ref{algo:pipeline2}), the current A/B test is deployed and the result of the A/B test is collected (line~\ref{algo:pipeline3}). The result of the A/B test is used to test the condition of the transition rules and to identify the next A/B test (lines~\ref{algo:pipeline4}-\ref{algo:pipeline5}). If the next A/B test is End, the execution of the A/B testing pipeline stops (as explained, we assume a consistent set of designed rules).
\vspace{5pt}\\\textbf{Conceptual architecture.} We present now the conceptual architecture of \approachPipelines{}. 
We start with 
the viewpoint of setting up and initiating a pipeline. Then we look at 
the viewpoint of the execution of a pipeline of A/B tests.

\paragraph{Architecture: Setting up and initiating a pipeline}

\begin{figure*}
    \centering
    \includegraphics[width=0.6\linewidth]{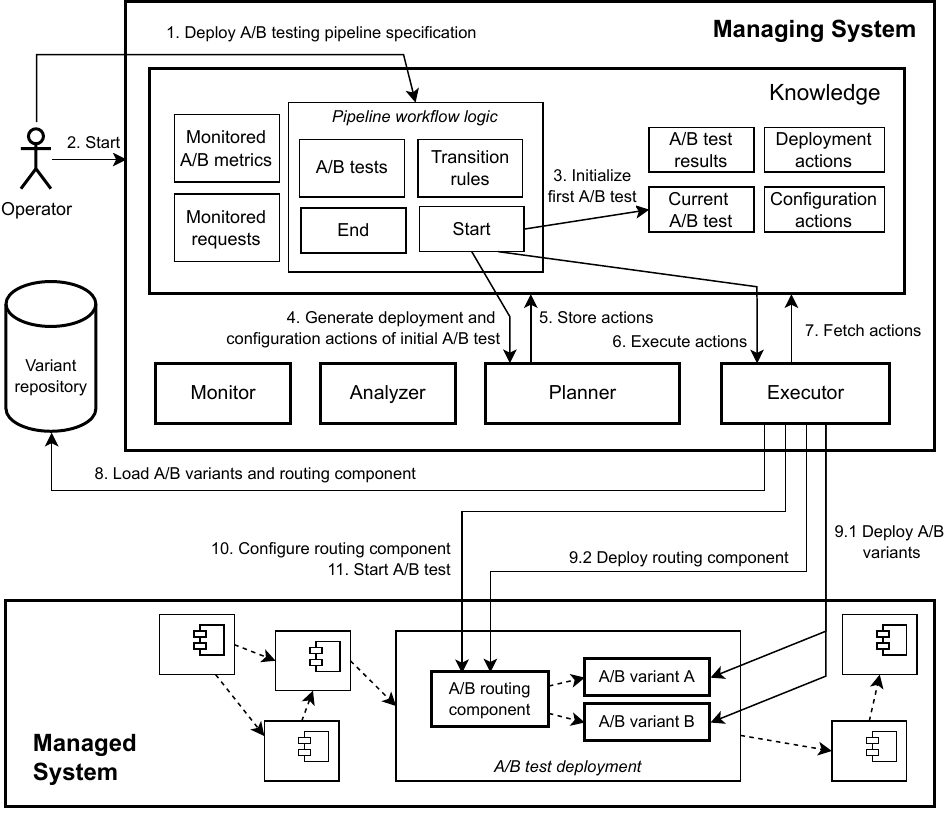}
    \caption{Architecture from the viewpoint of setting up and initiating an A/B testing pipeline}
    \label{fig:architecture-mape-pipeline-before}\vspace{-5pt}
\end{figure*}

Figure~\ref{fig:architecture-mape-pipeline-before} shows the architecture of \approachPipelines{} with a focus on setting up and initiating a pipeline of A/B tests. The process is initiated by an operator. After deploying the A/B testing pipeline specification at the pipeline workflow logic (1), the operator triggers the managing system to initiate the pipeline (2). The start component then initializes the current A/B test with the first A/B test of the pipeline (3). Next, start requests the planner to generate the deployment and configuration actions to deploy the A/B variants and the routing component for the first A/B test (4). The planner stores the actions in the knowledge repository (5). Start then triggers the executor to execute these actions (6). To that end, the executor fetches the actions (7) and loads the A/B variants and the routing component from the variant repository (8). Then the executor deploys the A/B variants (9.1) and the A/B routing component (9.2). Finally, the executor uses the configuration actions to configure the A/B routing component (10). This completes the setup and initialization of the A/B testing pipeline. The executor can then start the first A/B test of the pipeline (11). 

\paragraph{Architecture: Executing a pipeline}

\begin{figure*}
    \centering
    \includegraphics[width=.72\linewidth]{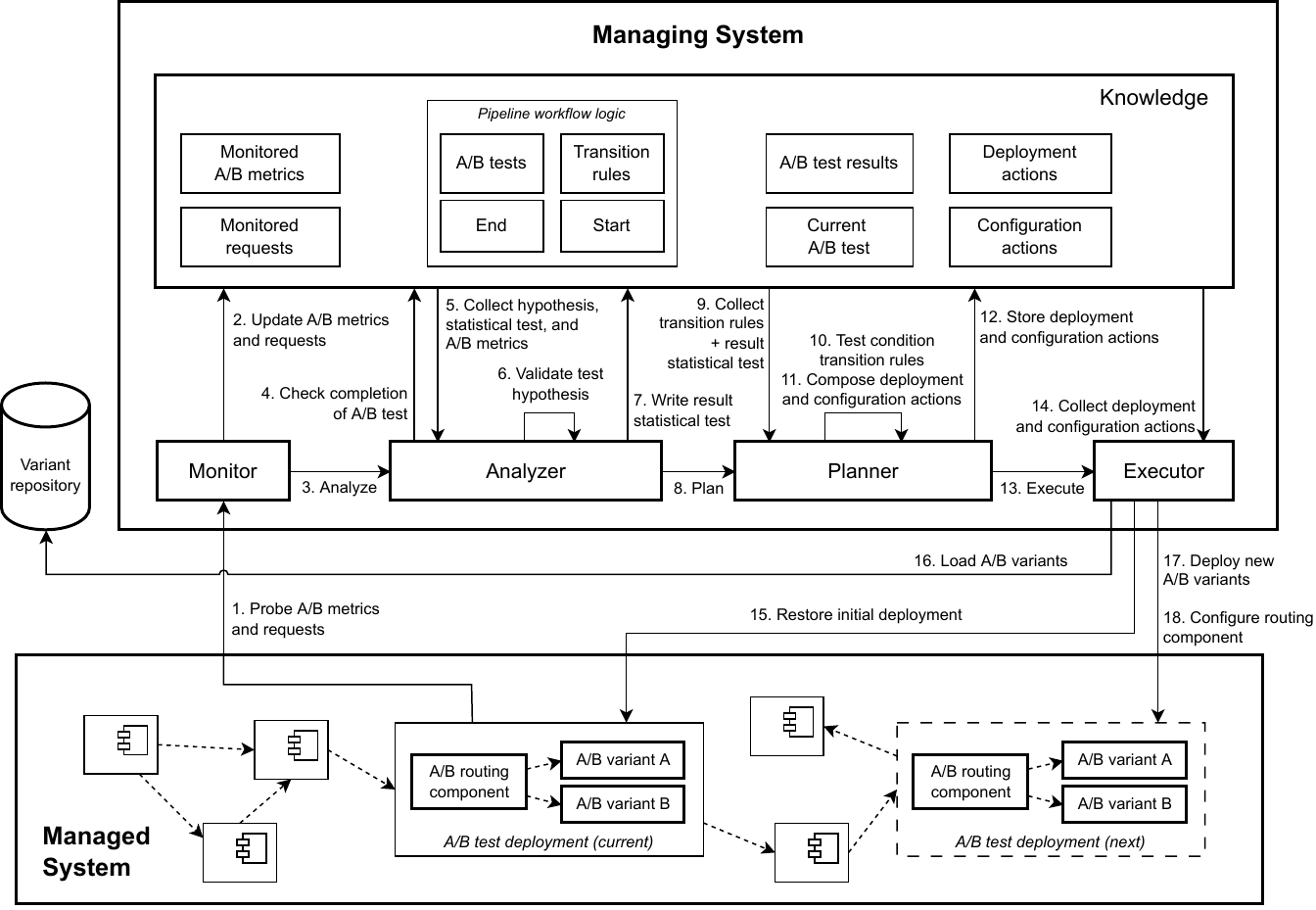}
    \caption{Architecture from the viewpoint of executing an A/B testing pipeline}
    \label{fig:architecture-mape-pipeline}\vspace{-10pt}
\end{figure*}

Figure~\ref{fig:architecture-mape-pipeline} shows the architecture of \approachPipelines{} with a focus on the execution of an A/B testing pipeline. We assume that the A/B test on the left-hand side is currently in execution and that this test uses the number of requests as the length of the experiment, while the A/B test on the right-hand side is the next A/B test. The monitor of the managing system starts with probing the status of the A/B metrics of the A/B test currently running and the number of requests invoked on the A/B variants (1); it then uses this data to update the knowledge repository of the managing system (2). Next, the monitor triggers the analyzer (3). The analyzer checks whether the number of invoked requests is sufficient to end the current A/B test (4). If this is the case, the process completes (not shown in Figure~\ref{fig:architecture-mape-pipeline}). Otherwise, the analyzer fetches the hypothesis, the statistical test, and the A/B metrics (5). The analyzer then tests the hypothesis (6), writes the result of the test to the knowledge repository (7), and triggers the planner (8).  
The planner collects the transition rules of the pipeline and the result of the statistical test (9) and tests the conditions of the transition rules (10). The planner then composes the deployment and configuration actions (11),  stores the actions in the knowledge repository (12), and triggers the executor (13). The executor collects the deployment and configuration actions (14). It then restores the initial deployment of the components involved in the current A/B test (15). Finally, if the next step in the pipeline is the execution of a new test, the A/B variants are loaded (16) and deployed (17) and the routing component is configured for the new A/B test (18). Otherwise, if the next step in the pipeline is the end of the execution, the stakeholders are informed that the execution of the pipeline has been completed and that the results of the A/B tests are available (not shown in Figure~\ref{fig:architecture-mape-pipeline}).

\subsection{\textcolor{new2}{Self-adaptation and Machine Learning to Split Populations}}

\approachPipelines{} supports a split of the population in an A/B testing pipeline according to predefined criteria. In this paper, we focus on segmenting a population based on properties or behaviors of users. This segmentation can then be used to target tailored A/B tests based on the appropriate property or type of user. 
\approachPipelines{} leverages \textcolor{new2}{self-adaptation and} machine learning techniques to support population splits. Segmenting the population offers an important benefit in terms of the efficiency of executing A/B testing pipelines: multiple parts of an A/B testing pipeline can be executed in parallel if the population segments assigned to each part of the pipeline are mutually exclusive, i.e., users are guaranteed to only be eligible for a single A/B testing pipeline.

\begin{figure}
    \centering
    \includegraphics[width=.8\linewidth]{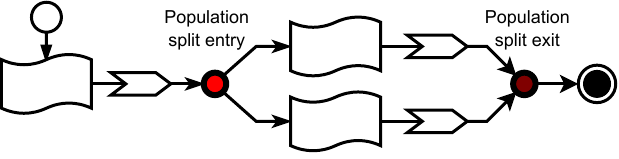}
    \caption{Visual notation population split in a simple pipeline}
    \label{fig:notation-visual-population-split2}\vspace{-10pt}
\end{figure}

We follow a similar structure as the section outlining the use of self-adaptation to automate the execution of pipelines to explain population splits in \approachPipelines{}. We present a specification to model a population split (tackling requirement R3). Then, we present the conceptual architecture of \approachPipelines{} focusing on population splits (tackling requirement R4).
\vspace{5pt}\\\textbf{Specification to model pipelines of A/B tests.}
To support modeling a population split, we present a simple specification. Figure~\ref{fig:notation-visual-population-split2} shows a visual notation of a population split used in a basic example of an A/B testing pipeline. 

We first explain the specification of a population split. Afterward, we extend the specification of an A/B testing pipeline with population splits. 

\paragraph{Specification population split} \mbox{}\vspace{5pt}\\
\indentSpec{}\textit{Pop-split = $<$ Pop-split-entry, Pop-split-exit $>$}\\\vspace{-5pt}

A population split (\textit{Pop-split}) consists of an entry (\textit{Pop-split-entry}) and an exit (\textit{Pop-split-exit}). We specify both parts now in detail, starting with the entry. \vspace{5pt}\\
\indentSpec{}\textit{Pop-split-entry = $<$ Split-prop, 
\{Sub-pipeline\},} \newline 
\indentSpecNewline{\textit{Pop-split-entry = $<$ }}\textit{\{Cond-stat\} $>$}\vspace{5pt}\\
\indentSpec{}\textit{Sub-pipeline = $<$ \{Subpl-ID, Start, \{AB-test\},} \newline
\indentSpecNewline{\textit{Sub-pipeline = $<$ }}\textit{\{Trans-rule\} $>$}\\\vspace{-10pt}\\

The split property (\textit{Split-prop}) defines the attribute on which the population is segmented. An example is the likelihood that a user makes a purchase on a website. \textcolor{new2}{In this example, machine learning could be used to predict the likelihood of making purchases based on the behavior of the user on the website.} The population is then segmented and assigned to a list of sub-pipelines (\textit{\{Sub-pipeline\}}) based on the split property. A sub-pipeline contains a unique identifier (\textit{Subpl-ID}). Additionally, it consists of a set of A/B tests (\textit{\{AB-test\}}) with a starting AB test (\textit{Start}) as the first test of the sub-pipeline, and a set of transition rules (\{Trans-rule\}). 
Since these sub-pipelines can be executed in parallel, they should not interfere \textcolor{new2}{to ensure the satisfaction of the SUTVA~\cite{primary-study-84, primary-study-21}.} (i.e., not involve shared components). Ensuring this constraint is the responsibility of the designer of the pipeline. The assignment of population segments to tests depends on the satisfaction of the specified set of conditional statements \textit{\{Cond-stat\}}, one per subsequent A/B test. 

The exit of a population split is defined as: \vspace{5pt}\\
\indentSpec{}\textit{Pop-split-exit = $<$ \{Assoc-trans-rule\}, Subseq-AB-test $>$}\\\vspace{-5pt}

An exit has an associated set of transition rules (\textit{Assoc-trans-rule}) that correspond to the completion of the different sub-pipelines determined by the population split.  After the exit of the population split, the A/B testing pipeline continues execution with its remaining part, i.e., the execution of the next A/B test or the execution ends (\textit{Subseq-AB-test}).

\paragraph{Updated specification A/B testing pipeline}

Lastly, we update the specification for an A/B testing pipelines that accommodates for population splits: \vspace{5pt}\\
\indentSpec{}\textit{AB-test-pl = $<$ \{AB-test\}, \{Trans-rule\}, \{Pop-split\},} \newline 
\indentSpecNewline{\textit{AB-test-pl = $<$ }}\textit{Start, End $>$}\\\vspace{-5pt}

An A/B testing pipeline comprises next to a set of A/B tests and transition rules also a set of population splits (\textit{\cbraces{Pop-split}}) that enable the enclosed sub-pipelines to run in parallel.

Algorithm~\ref{alg:application-population-split} describes the semantics of the application of a population split in A/B testing pipelines. We distinguish between the execution of a population split entry and a population split exit. 
The execution of a population split entry starts with instantiating a new knowledge component for each sub-pipeline (line~\ref{algo:split2}). In each knowledge instance, the population split adds a specific routing configuration according to the population split property and the conditional statement of the sub-pipeline (line~\ref{algo:split3}). Lastly, the population split entry starts the parallel execution of the sub-pipelines (line~\ref{algo:split4}). 
The execution of a population split exit removes the knowledge instances for each sub-pipeline from the knowledge repository (line~\ref{algo:split5}). Afterward, the A/B testing pipeline continues with the next activated A/B test in the pipeline (line~\ref{algo:split6}).

\begin{figure*}
    \centering
    \includegraphics[width=.85\linewidth]{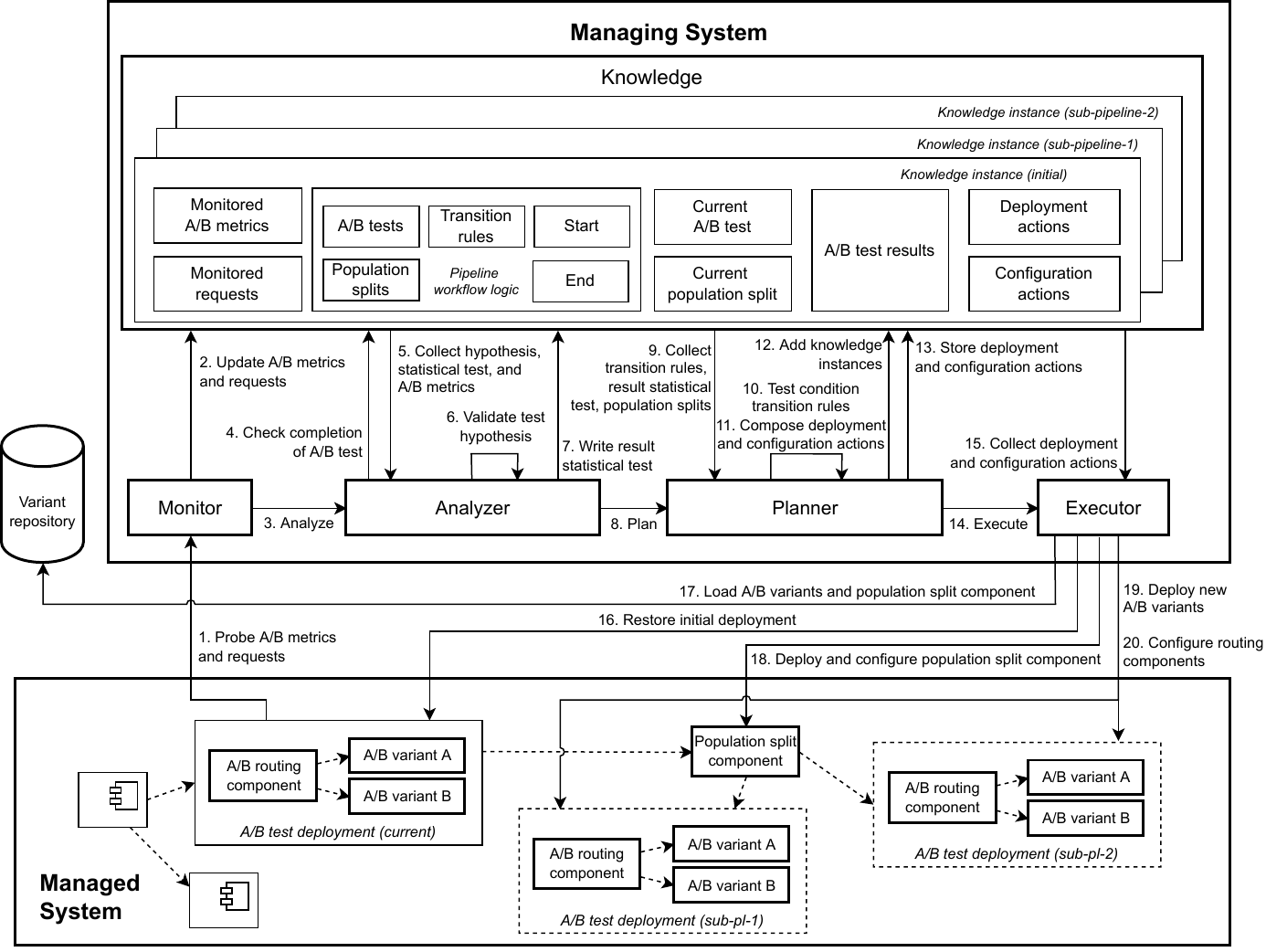}
    \caption{Architecture from the viewpoint of executing an A/B testing pipeline with population split}
    \label{fig:architecture-mape-pipeline-split}\vspace{-8pt}
\end{figure*}

\begin{algorithm}
\caption{Apply a population split.}\label{alg:application-population-split}
\begin{algorithmic}[1]
\Procedure{ExecuteSplitEntry}{$\text{split}$}
\State entry $\gets \text{split.Pop-split-entry}$
\For{$\text{sub-pl}, \text{cond}\ \text{\textbf{in}}\ (\text{entry.Sub-pipeline,} \newline \hspace*{15pt} \ \ \ \ \ \ \ \ \ \ \ \ \ \ \ \ \ \ \ \ \ \ \ \ \text{entry.Cond-stat})$} \label{algo:split1}
    \State k $\gets \text{Knowledge.addInstance(sub-pl.Subpl-ID)}$ \label{algo:split2}
    \State k.configureRouting(cond, entry.Split-prop) \label{algo:split3}
    
\EndFor
\State $\text{ExecutePipelines(entry.Sub-pipeline)}$ \label{algo:split4}
\EndProcedure
\State
\Procedure{ExecuteSplitExit}{\text{split}}
\State $\text{pipelines} \gets \text{split.Pop-split-entry.Sub-pipeline}$
\State $\text{ids} \gets \text{pipelines.Subpl-ID}$

\For{id \textbf{in} ids}
    \State Knowledge.removeInstance(id) \label{algo:split5}
\EndFor
\State Deploy(split.Pop-split-exit.Subseq-AB-test) \label{algo:split6}
\EndProcedure
\end{algorithmic}
\end{algorithm}
\noindent \vspace{-5pt}\\\textbf{Architecture with population split.}
Figure~\ref{fig:architecture-mape-pipeline-split} shows the architecture of \approachPipelines{} with a focus on population splits. We assume that the A/B test on the left in the managed system (denoted by \textit{A/B test deployment (current)}) is in operation. Steps 1 through 8 remain identical to the steps outlined in Figure~\ref{fig:architecture-mape-pipeline}. Once the planner is triggered, it collects the transition rules, results of the statistical test, and population splits from the knowledge (9). The planner then tests the conditions of the transition rules (10) and, if one of the transition rules is satisfied, composes the deployment and configuration actions (11). If the transition rule results in a regular A/B test, the flow as described in Figure~\ref{fig:architecture-mape-pipeline} continues. However, if a rule of a population split is satisfied, the planner prepares the knowledge and adds two knowledge instances; one for each sub-pipeline in the population split (12). Then, the planner stores the deployment and configuration actions in the knowledge (13), and triggers the executor (14). The executor collects the deployment and configuration actions from the knowledge (15). It then restores the initial deployment of the components involved with the A/B test (16). The executor then fetches the A/B variants for the sub-pipelines and the population split component (17). Next, it deploys and configures the population split component for the sub-pipelines (18). Finally, the executor deploys the new A/B variants for both sub-pipelines (19) and configures both routing components (20). The sub-pipelines can then start executing in parallel. 

\subsection{Concrete Realization of the Conceptual Architecture}

We implemented the conceptual architecture leveraging the SEAByTE~\cite{Quin2022} artifact that provides basic support for the automatic execution of pipelines of A/B tests applied to the domain of microservice-based systems (tackling R5). 
To implement the conceptual architecture of \approachPipelines{}, we extended the blueprints of experiments, transition rules, and pipelines in SEAByTE and added a blueprint for a population split. Then we extended the implementation of the managing system and we added a population split component to SEAByTE. 
We focus here on the realization of the population split. For further details, we refer to the SEAByTE website.\footnote{\url{https://people.cs.kuleuven.be/danny.weyns/software/SEAByTE/}}

\paragraph{Blueprint population split with machine learning}
Listing~\ref{listing:population-split} shows an example of the blueprint of a population split for the microservice-based application of SEAByTE.\vspace{5pt}

\begin{lstlisting}[language=JSON,
caption={Example blueprint of a population split.},
frame=single,
captionpos=b,
label={listing:population-split}]
{
  "name": "Population-split-purchases-prediction",
  "splitProperty": "purchase-likelihood",
  "pipelines": ["Review-pipeline", "Recommendation-pipeline"],
  "conditionalStatements": [{"==", 0}, {"==", 1}],
  "nextComponent": "end",
  "splitComponent": {
    "serviceName": "purchase-prediction-component",
    "imageName": "ml-purchase-filter"
  }
}
\end{lstlisting}

\noindent The elements in the blueprint of population splits are:

\begin{itemize}
    \item name: The name of the population split component.
    \item splitProperty: The property on which the population will be segmented.
    \item pipelines: The name of the pipelines to be started, in \textit{\cbraces{Sub-pipeline}} of a population split entry. In the example, we consider two sub-pipelines.
    \item conditionalStatements: The conditions that determine which segments of the population will take part in the designated A/B testing pipelines, corresponding to \textit{\cbraces{Cond-stat}} in the population split entry. In the example, users classified with a purchase likelihood of 0 will take part in the review pipeline, while 
    users with a 
    likelihood of 1 will take part in the recommendation pipeline.
    \item nextComponent: The component that follows after completing all pipelines in the population split. 
    \item splitComponent: The population split component that is responsible for exposing an API that classifies users on the provided split property. The population split component is deployed in docker with the given service name from the provided image name.
\end{itemize}

\begin{figure}
    \centering
    \includegraphics[width=0.85\linewidth]{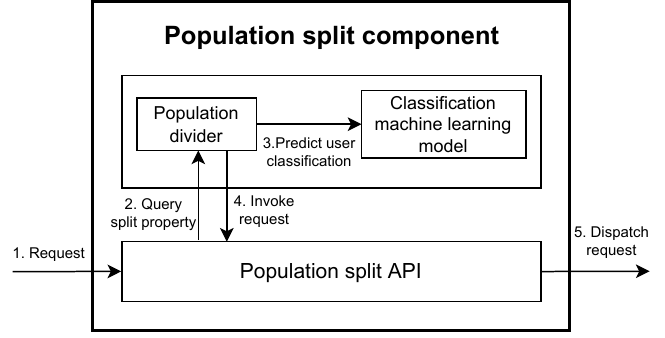}
    \caption{Architecture of the population split component}
    \label{fig:runtime-model-split-component}\vspace{-10pt}
\end{figure}

\paragraph{Realization of the population split component}
Figure~\ref{fig:runtime-model-split-component} shows the 
population split component supported by the enhanced version of SEAByTE during deployment. \textcolor{new2}{Before deployment in the live system (not shown in Figure~\ref{fig:runtime-model-split-component}), a classification machine learning model is loaded into the population split component. Prior to this, the model is trained using historically labeled user data. In our example, the feedback loop was responsible for keeping track of historical data in the application, and using this data to train the machine-learning model.}
At runtime, user requests are collected by the population split API (1). The API invokes a query with the split property to the population divider (2). The population divider then uses the trained classification machine learning model to predict the classification of the user (3). Next, the population divider invokes the request for the predicted class using the population split API (4) that then dispatches the request to the sub-pipeline of that class (5). 

\section{Evaluation}\label{sec:evaluation}

We start with presenting the results of a short survey with experts on the usefulness of the notation and infrastructure of \approachPipelines{}. Then we present the results of a series of tests that we performed to measure the gain obtained when using an automated A/B testing pipeline with a population split compared to a sequential pipeline. We also report the runtime overhead caused by a population split. For the tests, we use a scenario that we implemented in SEAByTE. We conclude the section with a discussion and analysis of threats to validity.

\subsection{Evaluation questions}

The evaluation aims at answering the following four  evaluation questions: 
\begin{description}
    \item[EQ1:] How do people knowledgeable in the topic appraise the \textcolor{new2}{usefulness} of the notation of \approachPipelines{} to model pipelines of A/B tests with population splits? 
    \item[EQ2:] How do people knowledgeable in the topic appraise the \textcolor{new2}{usefulness} of the infrastructure of \approachPipelines{} to execute pipelines of A/B tests with population splits?
    \item[EQ3:] What is the reduction in the number of requests that we can obtain to get statistically significant results of A/B tests in pipelines with population splits compared to sequential pipelines?
    \item[EQ4:] How much overhead do population splits introduce before and during executing A/B testing pipelines?
\end{description}

\subsection{Evaluation metrics}

To answer EQ1 and EQ2 we use a \textcolor{new}{single} questionnaire where experts could express their appraisal for \approachPipelines{} on a five-point Likert scale. To answer EQ3 and EQ4 we run experiments on a concrete system using the metrics of Table~\ref{tab:evaluation-metrics}.  The results  report the median from 15 runs. All evaluation materials with results are available on the SEAByTE website.

\begin{table}
    \centering
    \begin{tabular}{@{}m{1.1cm}m{1.25cm}m{5.35cm}@{}}
        \toprule
        \textbf{Question} & \textbf{Metric} & \textbf{Description} \\\midrule
        EQ3 & Reduction in requests & Difference in number of requests to obtain statistically significant results (from where $p \leq 0.05$) of an automatically executed pipeline of A/B tests  with and without population split. \\\midrule[0.25pt]
        EQ4 & Overhead & The time it takes (i) to train the machine learning model of the population split component, (ii) to deploy the population split component, (iii) to classify a request for a population split.\\\bottomrule
    \end{tabular}
    \caption{Metrics to answer the evaluation questions}
    \label{tab:evaluation-metrics}\vspace{-10pt}
\end{table}

\subsection{Evaluation Instruments and Settings}

\hspace{-10pt}\textbf{Population and questionnaire for answering EQ1 and EQ2.}
We invited 32 experts to participate in the questionnaire and received 19 valid answers. Fourteen answers (73.7\%) were from academics with practical experience and 5 answers (26.3\%) were from practitioners. The online questionnaire started with a brief introduction of \approachPipelines{}. Then we asked the following questions: 

\begin{enumerate}
    \item[Q1] How useful is automating A/B testing pipelines? 
    \item[Q2] How useful is a population split in A/B testing? 
    \item[Q3] How useful is the notation to specify A/B testing pipelines? 
    \item[Q4] How useful is an implementation that supports running A/B testing pipelines automatically?
\end{enumerate}

In addition for validity of the answers, we asked participants how familiar they are with self-adaptation and A/B testing.  \vspace{3pt}

\begin{figure}
    \centering
    \includegraphics[width=\linewidth]{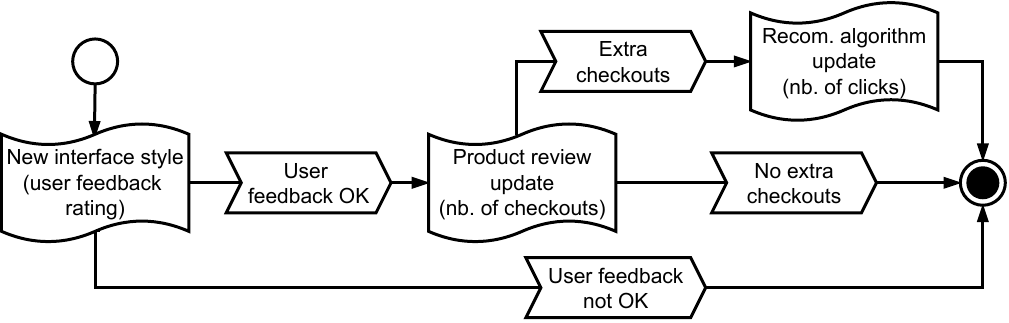}
    \caption{Pipeline for evaluating sequential runs of A/B tests}
    \label{fig:scenario-pipeline-sequential}\vspace{-10pt}
\end{figure}

\noindent \textbf{Scenarios for answering EQ3 and EQ4.} For the approach without population split we used the pipeline shown in Figure~\ref{fig:scenario-pipeline-sequential}.
The pipeline starts with an A/B test on a new style of a user interface of a web-store application. If the new style is favored by users, a new A/B test is launched that tests if users are more likely to purchase products when product reviews are presented at checkout, which may be an incentive for users to buy if they are hesitant to purchase at checkout.
If this leads to more checkouts, a new A/B test is launched that evaluates a new version of the recommendation algorithm with the aim of serving better-targeted recommendations. The hypothesis is that the new algorithm is more effective at generating recommendations that result in more purchases.

\begin{figure}
    \centering
    \includegraphics[width=\linewidth]{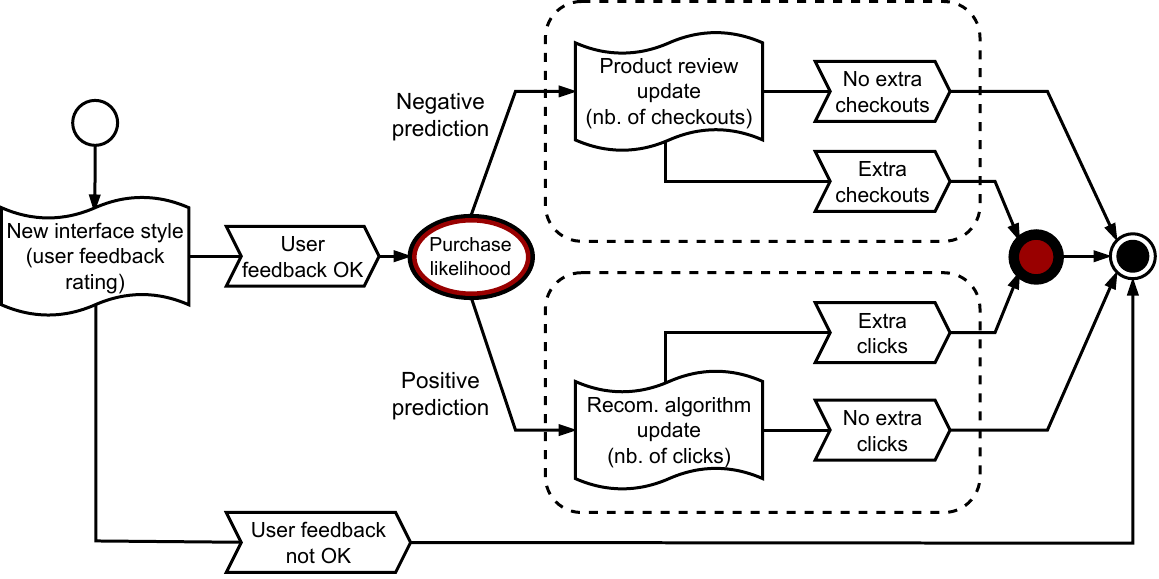}
    \caption{Pipeline for evaluating \approachPipelines{} with population split}
    \label{fig:scenario-pipeline-parallel}\vspace{-15pt}
\end{figure}

For \approachPipelines{} with a population split we used the pipeline shown in Figure~\ref{fig:scenario-pipeline-parallel}.
After successful completion of the A/B test of the new interface style (first A/B test of the pipeline), \textcolor{new2}{users are split between two pipelines according to their \emph{likelihood of purchasing} something in the web store.}
To that end, the population split uses a classifier to make predictions about the likelihood of a user making a purchase. Users that are predicted to make a purchase take part in the recommendation A/B test and the others take part in the review A/B test.
Since users belong to a single class, the review A/B test and the recommendation A/B test can run in parallel.  \vspace{3pt}

\noindent \textbf{Data sets.} We use 
publicly available datasets to (1) predict the likelihood of a user making a purchase in the web-store, and (2) to model user behavior in user profiles. 


\textit{Customer propensity to purchase}. The first dataset\footnote{\url{https://www.kaggle.com/datasets/benpowis/customer-propensity-to-purchase-data}} consists of $455,401$ labeled data samples, each with a user  identifier, 23 features describing actions taken by the user (e.g. requested information about a product) or characteristics of the user (e.g. the device used to visit the web-store), and a label about whether the user made a purchase in the web-store. Of all data samples, $4.2\%$ are labeled positively, i.e. a customer that made a purchase. We used 25\% of the data samples to train the machine learning model; the remaining 75\% was used to model the user behavior in the web store.

\textit{User profiles.} The second dataset\footnote{\url{https://www.kaggle.com/datasets/sergylog/ab-test-data}} contains $10,000$ data samples of an A/B test about the revenue obtained from users. In variant A $1.605\%$ of users make a purchase, while in variant B $1.435\%$ of users make a purchase. %
The third dataset\footnote{\url{https://www.kaggle.com/datasets/sergylog/ab-test-aggregated-data}} contains $120,000$ data samples with clicks of users. Of all samples, $14.70\%$ of users produce clicks in variant A, while $16.17\%$ of users produce clicks in variant B. We designed the user profile using these data sets.  \vspace{3pt}

\noindent \textbf{Test environment.} We implemented the pipelines in the extended version of SEAByTE. The tests were run on a machine with a 2 x Intel(R) Xeon(R) CPU E5-2680 v4 @ 2.40GHz 
processor and 16GB of RAM. 

For the population split component, we used a Stochastic Gradient Descent (SGD) classifier. 
\textcolor{new}{The classifier is trained by estimating the gradient of the loss for the training samples, and iteratively updating its model to minimize this loss.} In our evaluation, we employed an implementation of the classifier from the scikit-learn library~\cite{scikit-learn}.
The test setup and a replication package are available at the SEAByTE website. 

\textcolor{new}{For the evaluation of the approaches for sequential- and parallel A/B testing pipelines we used custom-developed solutions for A/B test execution (incl. user assignment, A/B metric tracking, and hypothesis testing). To the best of our knowledge, existing A/B testing tools such as Split.io, VWO, or Convert\footnote{\url{https://www.split.io/}, \url{https://vwo.com/}, \url{https://www.convert.com/}} do not offer support for implementing a population split component as described in our approach, hence restricting us from evaluating the parallel A/B pipeline in these tools. To ensure consistency in the execution of both pipelines, we also chose to execute the sequential A/B testing pipeline using our custom-developed solution, exploiting the same code for running, monitoring, and analyzing the A/B tests.}

\subsection{Evaluation Results}

\hspace{-10pt}\textbf{Survey Results (EQ1 and EQ2)} 
From the 19 valid answers\footnote{We removed four additional answers from participants that expressed that they have no basic knowledge of either self-adaptation or A/B testing.}, the average score for familiarity with self-adaptation was 4.11 and for familiarity with A/B testing was 3.21 on a Likert scale: 0 not familiar\,...\,5 an expert. These results show that the participants have the required knowledge to provide valid answers. 

We obtained an average score of 4.16 [$\pm$ 0.69] for the usefulness of automating A/B testing pipelines (Q1), while the score for the usefulness of population split was 4.21 [$\pm$ 0.85] (Q2) both on a Likert scale: 0 not useful ... 5 highly important. This underpins the importance of the research problem. 

\paragraph{EQ1: Usefulness notation} For the usefulness of the notation of \approachPipelines{} that supports modeling A/B testing pipelines with population splits (Q3), we obtained a score of 3.72 [$\pm$ 0.75]. This result shows that the participants appraise the usefulness of the notation provided by \approachPipelines{}.

\paragraph{EQ2: Usefulness infrastructure} For the usefulness of \approachPipelines{}'s infrastructure to run A/B testing pipelines with population splits (Q2) we obtained a score of 4.21 [$\pm$ 0.71]. These results show that the participants appraise the importance of the infrastructure provided by \approachPipelines{}. \vspace{5pt}

\hspace{-10pt}\textbf{Results of the Tests (EQ3 and EQ4)} 
We start with the results for the reduction in number of requests to get statistically significant results of A/B tests in pipelines with population splits compared to sequential pipelines. Then we look at the results for overhead caused by population splits.

\paragraph{EQ3: Reduction in the required number of requests with population splits} 
Since both A/B testing pipelines used in the evaluation (sequential Figure~\ref{fig:scenario-pipeline-sequential}, and parallel Figure~\ref{fig:scenario-pipeline-parallel}) start with a common A/B test that targets the whole population, the results of this A/B test are the same for both pipelines. Hence, we focus on the results of the two other A/B tests: the adjusted product review update A/B test and the adjusted recommendation algorithm A/B test. 
Table~\ref{tab:reduction-population-split} summarizes the results (median values of number of requests over 15 runs).

\begin{table}
    \centering
    \caption{Reduction number of required requests (EQ3)}
    \begin{tabular}{@{}llcc@{}}
        \toprule
        \textbf{Pipeline} 
        & \textbf{A/B test}& \begin{tabular}{@{}c@{}}\textbf{Number of}\\\textbf{requests (until}\\\textbf{$p \leq 0.05$)}\end{tabular} & \begin{tabular}{@{}c@{}}\textbf{Total requests}\\\textbf{required}\end{tabular}\\\midrule
        
        \multirow{3}{*}{Sequential} & \makecell{$S_1$ Recommendation\\[-2pt]\ \ \ \ \,update} & 112,000 & 112,000 \\
        & $S_2$ Review update & 27,000 & 27,000 \\\cmidrule[0.1pt]{2-4}
        & \multicolumn{2}{c}{\textbf{Total (SUM $S_1$ + $S_2$)}} & \textbf{139,000} \\\midrule[0.25pt]
        
        \multirow{3}{*}{Parallel} & \makecell{$S_1$ Recommendation\\[-2pt]\ \ \ \ \,update} & 1,000 & 24,038 \\
         & $S_2$ Review update & 26,000 & 27,128 \\\cmidrule[0.1pt]{2-4}
        & \multicolumn{2}{c}{\textbf{Total (MAX $P_1$, $P_2$)}} & \textbf{27,128} \\\bottomrule
    \end{tabular}\vspace{-10pt}
    \label{tab:reduction-population-split}
\end{table}

For the review update A/B test, we observe a small difference in favor of the test with the population split. 
The review update A/B test in the sequential pipeline obtains a statistically significant result at $27,000$ requests, versus $26,000$ requests for the pipeline with population split.  
The results show that the population split component assigned $95.84$\% of the requests to the review update A/B test. The reduction in the number of requests to finish the A/B test with population split is $3.70$\%.

\begin{figure}
    \centering
    \includegraphics[width=\linewidth]{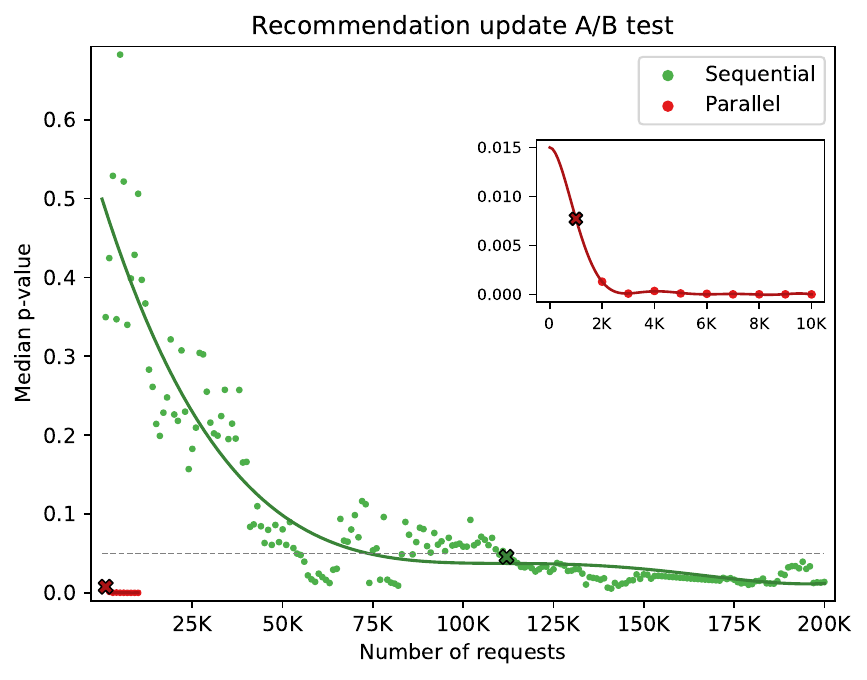}
    \caption[]{Median p-values in the recommendation update A/B test for the sequential and parallel pipelines (detail top right)\footnotemark}
    \label{fig:p-value-progress-both}\vspace{-10pt}
\end{figure}

For the recommendation update A/B test, we observe a large improvement in favor of the population split. Sequential execution obtains a statistically significant result after $112,000$ requests. Parallel execution immediately reaches statistical significance after $1,000$ requests. The population split component assigned $4.16$\% of the requests to the recommendation update A/B test. The reduction in the number of requests for the recommendation update A/B test is $99.11$\%. This very high number shows that the machine learning component is able to separate the two classes of users very well. The requests invoked on the recommendation update A/B test ensure that the test quickly obtains a statistically significant result.  
Figure~\ref{fig:p-value-progress-both} illustrates the progress of the p-values for the experiment of the recommendation update A/B test.

\begin{figure}
    \centering
    \includegraphics[width=\linewidth]{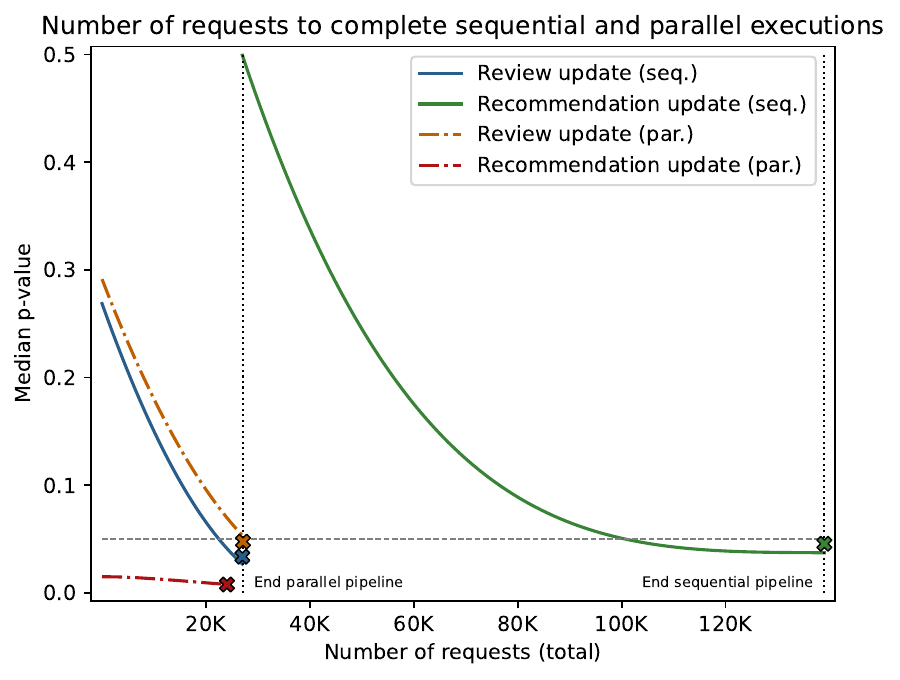}
    \caption[]{Progress of sequential and parallel execution for both the review update and recommendation update A/B tests\textsuperscript{\ref{footnote:lines}}}
    \label{fig:combined-review-recommendation}\vspace{-10pt}
\end{figure}

\footnotetext{\label{footnote:lines}The regression lines denote a polynomial fit over the data.}
Lastly, we look at the total number of requests required to complete the two tests with sequential and parallel execution. The execution with the sequential pipeline requires a total of $139,000$ requests to finish the execution of the two A/B tests with statistically significant results, i.e., the sum of $27,000$ and $112,000$ for the review update and recommendation A/B test, respectively. 
The execution of the pipeline with the population
split finished after 27,128 requests in total, i.e., the total number of requests required to obtain statistically relevant results (26,000 and 1,000). 
Figure~\ref{fig:combined-review-recommendation} illustrates the progress of the p-values over the requests to complete the tests.
The overall gain in required requests 
with population split is $80.48$\%. We conclude that splitting a population based on the specific behavior of users realizes a significant improvement in required requests to complete the tests with significant results. 

\paragraph{EQ4: Overhead introduced by population splits} 
Table~\ref{tab:overhead-population-split} provides an overview of the overhead introduced by a population split during preparation and operation.

\begin{table}
    \centering
    \caption{Overhead population split component (EQ4)}
    \begin{tabular}{@{}llc@{}}
        \toprule
        \textbf{Overhead type} 
        & \textbf{Timing}& \begin{tabular}{@{}c@{}}\textbf{Overhead}\\\textbf{(average, in msec)}\end{tabular} \\\midrule
        Training classifier & Offline & 324 \\
        Component deployment & Deployment & 6433 \\
        Prediction user class & Runtime & 0.3 \\\bottomrule
    \end{tabular}\vspace{-15pt}
    \label{tab:overhead-population-split}
\end{table}

Preparing the population split component consists of two steps: training the learning model and deploying the component. To train the machine learning model, historically labeled data is used. In the evaluation setting, this data was derived from the data sets we used. Training the machine learning model took on average 324 ms. The deployment time of the population split component heavily depends on its implementation. In the case of SEAByTE, we use Docker to create and start a container that contains the population split component as a micro-service.
The creation and startup took on average 1433 ms. %
In addition, Docker waits 5 sec (fixed) to check that the container is healthy. This resulted in a total average deployment time of 6433 msec. This overhead is not relevant \mbox{compared to the time it takes to run A/B tests in practice.}

During operation, the population split component classifies the population (e.g., purchasing or non-purchasing) using the trained classifier model.
Predicting the class of a user took less than a millisecond (0.3 msec). This time is also negligible compared to the time it takes to run A/B tests in practice. 

\subsection{Discussion and Threats to Validity}

\hspace{-10pt}\textbf{Discussion} 
We start with critical reflections on the tests.  
    
    The significant increase in efficiency with a population split results from i) the ability to run the review and recommendation updates in parallel, and ii) the population split allows for targeted A/B testing to relevant segments of the population. \color{black}
    
    The performance and accuracy of the machine learning model affect the result of the method: a model that makes bad predictions will divide a population wrongly thus affecting the results of the A/B tests. The model used in the evaluation performed very well, demonstrating the benefits of population splits for targeted A/B tests.
    
    The introduction of a population split introduces additional latency to the processing of requests. This extra latency can be detrimental to the user experience as noted by practitioners at Booking.com~\cite{Bernardi2019}. Hence, the designer needs to ensure that the time the machine learning model takes to produce predictions is acceptable to the users.
    
    \color{new2} Besides the evaluation scenario of this paper, the population split component could also be used to detect undesirable outcomes of A/B tests early in specific population segments. In case an A/B test on a population segment produces regressive results, practitioners can specify that the A/B pipeline should shut down prematurely. Otherwise, A/B testing can continue on the other population segments, if desired. \color{black}
    
    
    \color{new2} Creating a labeled dataset to train the classifier can also carry a substantial cost. We leave delving into this topic more comprehensively for future work. \color{black}
    
    In the evaluation, we split the population into binary segments. However, \approachPipelines{} does not impose this limitation. 
    Future work could explore dividing the population into more than two segments, or explore the use of unsupervised learning to split the population to avoid the need for labeled datasets. \color{black}


\vspace{5pt}
\noindent \textbf{Threats to validity} 
We discuss construct and external validity threats of both the survey and the tests. 

\paragraph{Construct validity} 

The questionnaire probed the usefulness of automating A/B testing pipelines and population split in general and the support offered by \approachPipelines{} in particular. Since we used closed questions, the participants were not able to provide nuances in their answers. Moreover, we provided only a brief introduction to \approachPipelines{}, so the participants may not fully grasp the usefulness of the notation and infrastructure. We acknowledge that the validity of the small survey may be limited. However, we believe that the results provide a first good indication. To obtain deeper insight, additional studies are needed where participants effectively use the notation and infrastructure.   

To measure the reduction in requests in the tests, we defined statistical significance from the point where $p \leq 0.05$ of the experiments of A/B tests. We used the median values over 15 runs to account for stochasticity in the data. The number 15 was empirically determined and may differ for other settings.


\paragraph{Internal validity}
\textcolor{new2}{To evaluate \approachPipelines{}, we used publicly available datasets to model the behavior of users in the web store. However, limited information is available about the origin of the datasets (as mentioned by others~\cite{primary-study-1623}), raising a threat to the internal validity of the results. To fully mitigate this threat, an industrial case study should be conducted.}

\paragraph{External validity}


The questionnaire only involved 19 participants with mixed knowledge and experience in self-adaptation and A/B testing. 
A more extensive survey and more participants would enhance the generalization of the results.

Since we only evaluated the approach for one concrete scenario in the context of a web store, we cannot make general claims about the applicability of the approach in different contexts. We anticipate that the technology and domain used for the evaluation are particularly relevant for contemporary software systems. In addition, we used external data sets to avoid bias. Lastly, we also provide a replication package~\cite{artifact-website} that is available for other researchers to replicate the results.

\color{new2} To answer EQ4, we measured (1) the time it took to train the machine learning model used in the population split component and (2) the time it took to deploy the component and use the component at runtime. We acknowledge that these measurements depend on the algorithms and technology used. \color{black}

\section{Conclusion and Future Work}\label{sec:conclusion}

Leveraging self-adaptation and machine learning, we presented \approachPipelines{}, a new approach to automating the execution of pipelines of A/B tests with 
support for splitting populations. We specified the elements of \approachPipelines{} and based on that presented a conceptual architecture. We instantiated this architecture extending the SEAByTE artifact. A small survey underpins the relevance of the approach and its usefulness. Test results on a realistic micro-service application show that \approachPipelines{} 
accelerates the identification of statistically significant results of the A/B tests in the required  number of requests with $80.48$\% compared to traditional sequential tests on the general population. In future work, we plan to study the identification of user groups without having access to labeled data leveraging unsupervised learning. This opens possibilities of automatically setting up A/B tests by experimenting with the target group of the A/B tests, without explicitly specifying the complete A/B tests in the pipeline. \textcolor{new}{We also plan to investigate ways of incorporating and potentially altering the A/A testing process in the approach.} Lastly, we aim to provide a full-fledged tool for specifying A/B testing pipelines with \approachPipelines{} along with a framework to execute the pipelines that can be tailored to the domain and needs at hand.

\bibliographystyle{ieeetr}
\bibliography{main}

\end{document}